\documentclass[aps,prx,superscriptaddress,twocolumn,showpacs,intlimits,amsmath,amssymb,floats]{revtex4-1}

\usepackage{times}
\usepackage{epsfig}
\usepackage{graphicx}
\usepackage{pstricks}
\usepackage{dcolumn}		
\usepackage{bm}			
\usepackage{ulem}
\usepackage{units}		
\usepackage{color}

\newcommand{\kw}[1]{\textcolor{red}{#1}}

\begin{document}

\title{Directly characterizing the relative strength and momentum dependence of electron-phonon coupling using resonant inelastic x-ray scattering} 
\author{T. P. Devereaux}
\affiliation{Stanford Institute for Materials and Energy Sciences, SLAC National Accelerator Laboratory, 2575 Sand Hill Road, Menlo Park, CA 94025.}
\affiliation{Geballe Laboratory for Advanced
Materials, Stanford University, CA 94305.}
\author{A. M. Shvaika}
\affiliation{Institute for Condensed Matter Physics of the National Academy of Sciences of Ukraine, Lviv, 79011 Ukraine.}
\author{K. Wu}
\affiliation{Stanford Institute for Materials and Energy Sciences, SLAC National Accelerator Laboratory, 2575 Sand Hill Road, Menlo Park, CA 94025.}
\author{K. Wohlfeld}
\affiliation{Institute of Theoretical Physics, Faculty of Physics, University of Warsaw, Pasteura 5, PL-02093 Warsaw, Poland}
\author{C. J. Jia}
\affiliation{Stanford Institute for Materials and Energy Sciences, SLAC National Accelerator Laboratory, 2575 Sand Hill Road, Menlo Park, CA 94025.}
\author{Y. Wang}
\affiliation{Stanford Institute for Materials and Energy Sciences, SLAC National Accelerator Laboratory, 2575 Sand Hill Road, Menlo Park, CA 94025.}
\author{B. Moritz}
\affiliation{Stanford Institute for Materials and Energy Sciences, SLAC National Accelerator Laboratory, 2575 Sand Hill Road, Menlo Park, CA 94025.}
\author{L. Chaix}
\affiliation{Stanford Institute for Materials and Energy Sciences, SLAC National Accelerator Laboratory, 2575 Sand Hill Road, Menlo Park, CA 94025.}
\author{W.-S. Lee}
\affiliation{Stanford Institute for Materials and Energy Sciences, SLAC National Accelerator Laboratory, 2575 Sand Hill Road, Menlo Park, CA 94025.}
\author{Z.-X. Shen}
\affiliation{Stanford Institute for Materials and Energy Sciences, SLAC National Accelerator Laboratory, 2575 Sand Hill Road, Menlo Park, CA 94025.}
\affiliation{Geballe Laboratory for Advanced
Materials, Stanford University, CA 94305.}
\affiliation{Dept. of Physics and Applied Physics, Stanford University, CA 94305.}
\author{G. Ghiringhelli}
\affiliation{CNR-SPIN and Dipartimento di Fisica, Politecnico di Milano, I-20133 Milano, Italy.}
\author{L. Braicovich}
\affiliation{CNR-SPIN and Dipartimento di Fisica, Politecnico di Milano, I-20133 Milano, Italy.}
\date{\today}

\begin{abstract}
The coupling between lattice and charge degrees of freedom in condensed matter materials is ubiquitous and can often result in interesting properties and ordered phases, including conventional
superconductivity, charge density wave order, and metal-insulator transitions. Angle-resolved photoemission spectroscopy and both neutron and non-resonant x-ray scattering serve as effective probes for determining the behavior of appropriate, individual degrees of freedom -- the electronic structure and lattice excitation, or phonon dispersion, respectively. However, each provides less direct
information about the
mutual coupling between the degrees of freedom, usual through self-energy effects, which tend to renormalize and broaden spectral features precisely where the coupling is strong, impacting ones
ability to quantitively characterize the coupling. Here we demonstrate that resonant inelastic x-ray scattering, or RIXS, can be an effective tool to directly determine the relative strength and momentum
dependence of the electron-phonon
coupling in condensed matter
systems. Using a diagrammatic
approach for an 8-band model
of copper oxides, we study the
contributions from the lowest
order diagrams to the full
RIXS intensity for a
realistic scattering geometry,
accounting for matrix element
effects in the scattering
cross-section as well as the
momentum dependence
of the electron-phonon coupling
vertex. A detailed examination of
these maps offers a unique perspective
into the characteristics of electron-phonon
coupling, which complements
both neutron and non-resonant
x-ray scattering, as well as
Raman and infrared conductivity.
\end{abstract}
\maketitle

\section{Introduction}

The ability to characterize excited states and fundamental excitations in solids remains one of the forefront challenges in condensed matter physics. This is particular evident in materials that display emergent phases, such as unconventional superconductivity, that stem from intertwined lattice, charge, spin and orbital degrees of freedom \cite{Fradkin2015}. Typically simple pictures for excitations derived from well-controlled perturbation expansions do not capture the richness of these materials due to the lack of a small parameter around which to expand. This is also evident particularly for excitations around a quantum critical point in correlated materials, where the low energy degrees of freedom are presumably intertwined. The lack of knowledge of how to truly quantify the strength of couplings for all momenta throughout the Brilloiun zone (BZ) has been one of the major roadblocks to understanding the complex phase diagrams that typically emerges near phase boundaries between various electronic, magnetic, or lattice instabilities.

In the last decade resonant inelastic x-ray scattering (RIXS) has enabled seminal progress in the understanding of fundamental excitations in correlated materials \cite{Ament}. 
Exploiting the role of strong spin-orbit coupling in $2p$ core levels in Cu $L$-edge RIXS, it has been shown recently that spin flip excitations can be probed across a wide variety of materials, complementing RIXS studies of charge transfer, $d-d$ orbital, and bi-magnon excitations down to a now available resolution better than 100 meV \cite{GG,AS}. Perhaps more intriguingly from
the perspective of a single tool that can provide complete characterization of charge, orbital, spin, and lattice exciations, excitations off of the elastic line oxygen $K$-edge that occur at multiples of optical phonon energies have become visible in one dimensional edge-shared CuO systems \cite{LeePhononsI}, while phonon side bands have emerged off of $d-d$ excitations measured at the Cu $L$-edge \cite{LeePhononsII}. 
This progress have been made with a resolution of about 120 meV at the Cu $L$-edge obtained already ten years ago at the Swiss Light Source (SLS) \cite{Ghiringhelli2006,AS}. The possibilities expanded further less than one year ago by the new standard of 35 meV at the Cu $L$-edge at the European Synchrotron Radiation Facility (ESRF) \cite{Brookes2016}. Since the phenomena of emergence involves necessarily the intertwined nature of different degrees of freedom in the excitation spectra, knowledge of the changes of excitation spectra is a goal that is well-addressed in present and future RIXS experimental configurations.

Specifically, one of the outstanding issues in condensed matter physics is to determine the strength of coupling of electrons to lattice excitations across the entire BZ. This has important ramifications to our understanding of emergent phenomena such as superconductivity and density wave order. For conventional isotropic superconductors, the momentum dependence of the electron-phonon coupling is rather irrelevant to pairing as all modes contribute to the Cooper instability. However, this is not the case in unconventional (sign-changing) anisotropic superconductors \cite{Bulut,OKA,Johnston}. While electron-phonon coupling at large momentum transfers that connects momentum points with different signs of the superconducting gap leads to a suppression of superconductivity, scattering that involves small momentum transfers always enhances superconducting pairing \cite{JJLee2014}. Likewise, it has now become accepted that for dimensions greater than 1, the momentum dependence of the electron-phonon coupling $g({\bf q})$ largely controls the ordering wavevector \cite{CDW}. Therefore a determination of $g({\bf q})$ is highly desirable on rather general grounds.

\begin{figure}
\includegraphics[width=0.8\columnwidth]{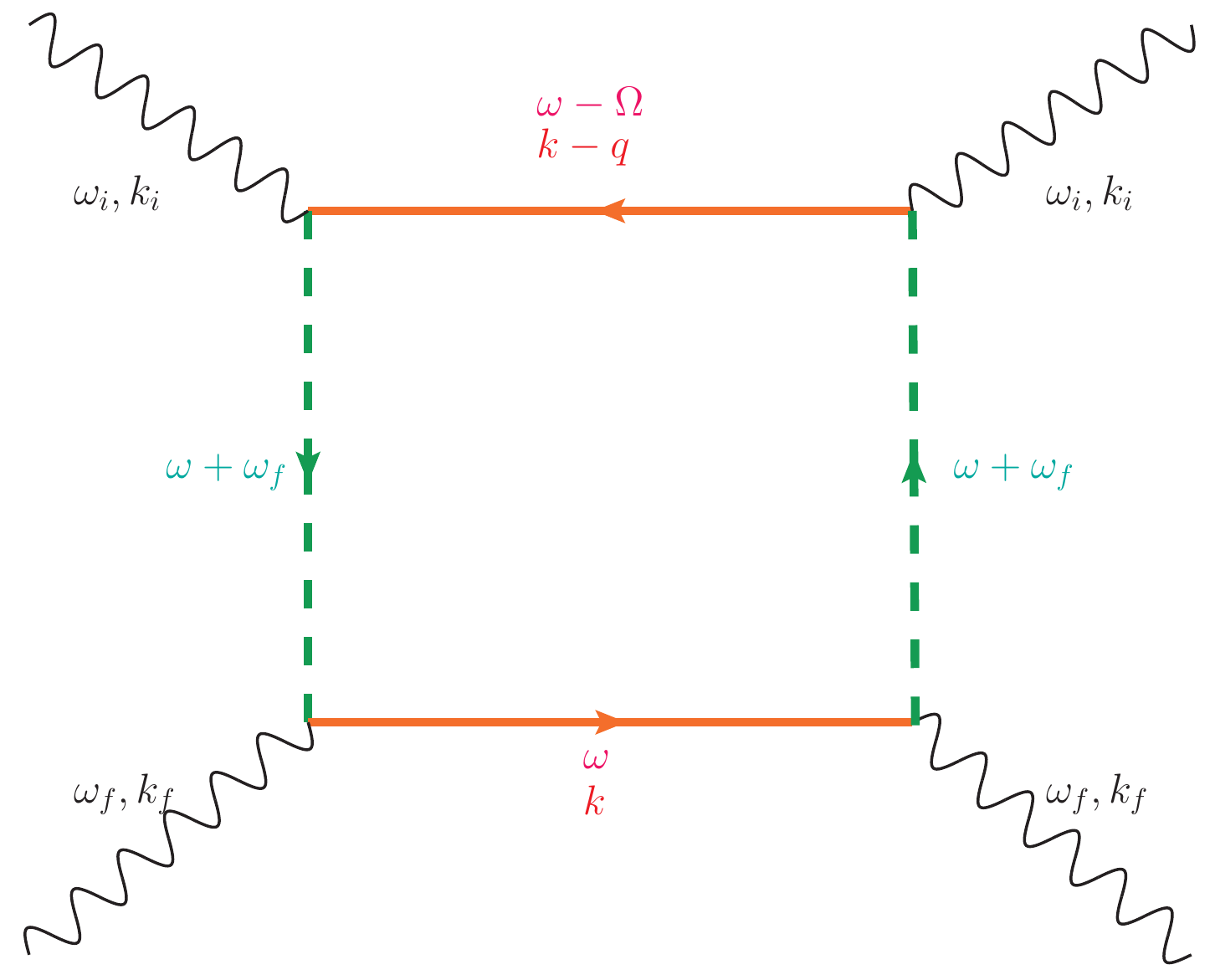}
\caption{\label{fig:bare}  
{Electron-hole contribution to RIXS (bare diagram). The dotted lines denote the Cu $2p$ hole and solid lines the $3d$ conduction electrons. Notation for the vertices have been suppressed.}}
\end{figure}

While Raman and optical measurements can extract the coupling $\lambda$ at zone center \cite{buckling,RamanRMP,neutron}, a reliable method for extracting the strength of lattice couplings for all momenta to date has been inferred indirectly from angle-resolved photoemission (ARPES) experiments, and more directly from inelastic neutron or high-energy non-resonant x-ray scattering, which couple directly to lattice vibrations \cite{Johnston,SpecRev,AREV,IXS,Reznik,BiPhonons,SJJacoustic}. While ARPES probes electrons at well defined electron momenta in the BZ, a sum over phonons at all bosonic momenta contributes to renormalization effects detected as "kinks" in dispersion, or abrupt changes of spectral linewidths at the energies of the lattice modes. This precludes as a practical manner a momentum-resolved way to detect $g({\bf q})$ throughout the BZ. 

On the other hand momentum dependent coupling can be determined from lineshape fitting of x-ray or neutron scattering. However, these techniques present a fundamental challenge to the extraction of lattice coupling itself. Due to the scattering process which couples to the atomic degrees of freedom, lattice vibrations are particular sharp and easy to detect when the lattice coupling to electrons is weak. In contrast, when the coupling is large, as seen for example in bond-stretching modes in the cuprates or near charge density wave instabilities, phonon lineshapes become broad in both energy and momentum which interfere with other phonons\cite{Reznik}. Thus the region of energy and momentum where the lattice coupling is strongest tends to be also the region where a direct extraction of the coupling is most difficult. Therefore one desires a complementary method that can be used to determine coupling in this region.

\begin{figure}
\includegraphics[width=0.8\columnwidth]{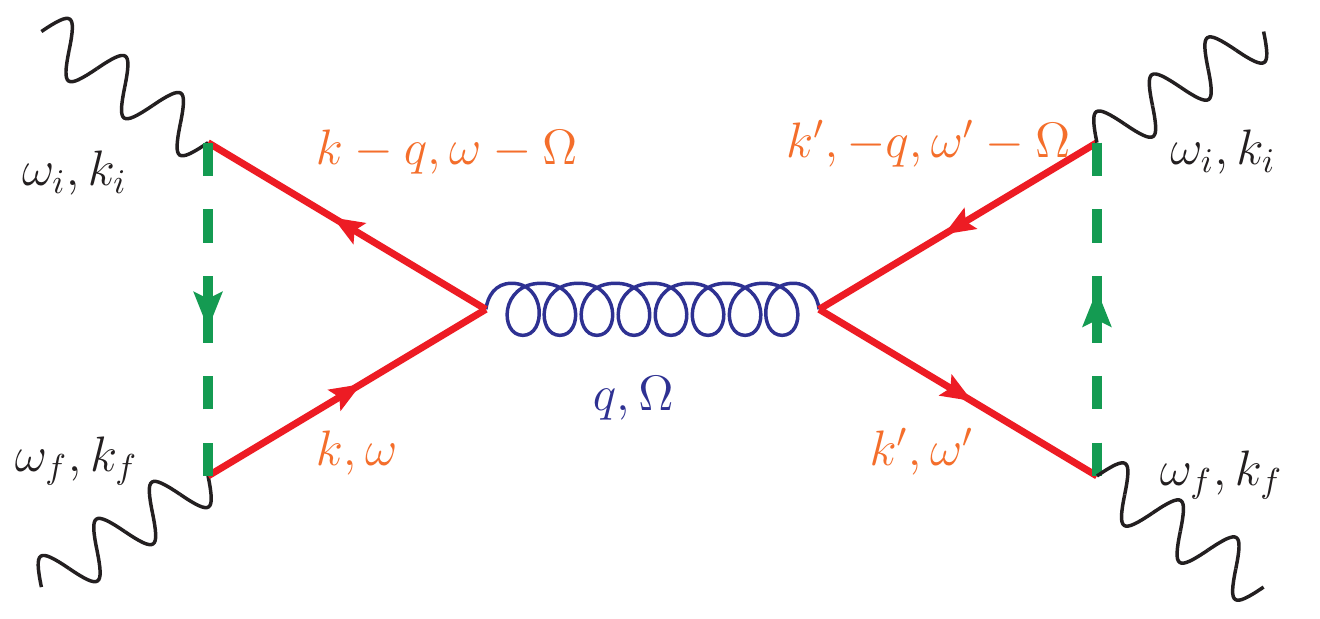}
\caption{\label{fig:Phonon}  
{Leading order one-phonon contribution to RIXS. The dotted lines denote the Cu $2p$ hole and solid lines the $3d$ conduction electrons, and the circular line denotes the phonon. Notation for the vertices have been suppressed.}}
\end{figure}

As suggested in earlier studies \cite{LeePhononsI,exprixs,EPL,Yao}, RIXS offers a unique insight into the momentum dependence and magnitude of electron-phonon coupling. In this paper we examine this selectivity of the light scattering process to determine how and which phonons can be resonantly excited during a resonant x-ray process, and the resulting information that can be obtained about the relative strength and momentum dependence of electron-phonon coupling. We primarily consider an $L$-like resonant x-ray scattering where a core electron is excited into the conduction band, and assume that the most interesting phonons are those that couple to the conduction/valence electrons/holes. We will neglect effects from direct phonon coupling to core levels, which would primarily only affect the intermediate state (absorption) profile. We particularly highlight the role of polarization and selection rules of the light-electron-phonon coupling, which are already well-characterized only at zone center in Raman measurements. Specifically we find that RIXS can provide detailed clues in regions of momentum space with strong electron-phonon coupling with atomic specificity, to further complement neutron and IXS measurements.

\section{Expressions for phonon contributions to RIXS}

Since we are interested in the generic ways in which phonons can be coupled in RIXS, we consider weak electron-phonon coupling and examine the one-phonon contribution to RIXS (Figs. \ref{fig:bare} and \ref{fig:Phonon}). While other considerations involving the ratio of main intensities to satellites offer a way to quantify electron-phonon coupling for generic model Hamiltonians \cite{LeePhononsI,EPL}, our approach follows an extension of resonant Raman scattering from phonons, extended to arbitrary momentum transfers and different x-ray resonant edge processes. We examine the interplay of light polarization configurations with specific momentum dependent electron-phonon coupling for different classes of coupling (for example, deformation, piezoelectric, and electrostatic), for some oxygen modes common to perovskites. Here we neglect the role of Coulomb interactions, both among and between the valence electrons and the core hole. While the role of Coulomb interactions is crucial for indirect RIXS processes \cite{Ament}, and for couplings at BZ center, for larger momentum processes it plays only a minor role \cite{Johnston,tpd,multiband}. 

The phonon branches are characterized by a frequency $\Omega_{\nu}({\bf q})$ for each mode $\nu$ as well as a generalized coupling to electrons $g^{\alpha,\beta}_{\nu}({\bf k,q})$. Here $g^{\alpha,\beta}_{\nu}({\bf k,q})$ denotes the coupling of phonon mode $\nu$ to an electron in band $\alpha$ carrying momentum ${\bf k}$ scattering into band $\beta$ with momentum ${\bf k+q}$. The total RIXS response contains two contributions from these two diagrams $ \chi  = \chi_{\text{bare}}+ \chi_{\text{phonon}},$
each of which depend upon $\mathbf{k_i},\mathbf{k_f},\mathbf{q},\omega_i,\omega_f,\Omega$
where $\mathbf{q=k_i-k_f}$ and $\qquad \Omega = \omega_i - \omega_f $.

\begin{widetext}
For the bare loop contribution (Fig. \ref{fig:bare}) we have
\begin{align}
\chi_{\text{bare}}^{\mu,\nu}(\mathbf{k_i},\mathbf{k_f},\mathbf{q}|\omega_i,\omega_f,\Omega) = \frac{1}{N} \sum_{\mathbf{p}} \left[\mathbf{e_i}\cdot\mathbf{d}_{\mu,\nu}(\mathbf{p+q,p-k_f})\right] \left[\mathbf{e_f}\cdot\mathbf{d}_{\nu,\mu}(\mathbf{p-k_f,p})\right]
\nonumber\\ 
\times\left[\mathbf{e_i}\cdot\mathbf{d}_{\mu,\nu}(\mathbf{p-k_f,p+q})\right] \left[\mathbf{e_f}\cdot\mathbf{d}_{\nu,\mu}(\mathbf{p,p-k_f})\right]
\nonumber\\ 
\times\int_{-\infty}^{+\infty} d\omega \left[f(-\omega-\Omega)-f(-\omega)\right] \left|G_{\nu}(\mathbf{p-k_f},\omega-\omega_f)\right|^2 
\nonumber\\ 
\times\left[-\frac{1}{\pi} \Im G_{\mu}(\mathbf{p},\omega)\right] \left[-\frac{1}{\pi} \Im G_{\mu}(\mathbf{p+q},\omega+\Omega)\right],
\end{align}
\end{widetext}
where $\Im$ denotes the imaginary part and $\mu,\nu$ denote the conduction, core electron, respectively, appropriate for a direct RIXS transition.
The matrix elements depend generally on the incident and scattered photon momentum $({\bf k_{i,f}})$, energy $(\omega_{i,f})$ and polarizations ${\bf e_{i,f}}$, respectively, including the light dipole couplings  $\mathbf{d}_{\mu,\nu}(\mathbf{k}, \mathbf{q})$ involving the resonant photoexcitation of a $\nu$ core hole with momentum $\mathbf{q}$ into the $\mu$ band with momentum $\mathbf{k}$. 


As a concrete example, our target will be an understanding of lattice coupling in the cuprates, to which end we consider an Cu $L-$edge scattering process where a Cu $2p$ core electron is photoexcited into the $3d$ valence band at a resonant energy around 931 eV (Cu $L_3$-edge). However, the formalism will work equally well for other resonant excitations involving core electrons and valence states, such as the oxygen $K$-edge.

We only consider particle-hole excitations in the Cu-O hybridized band which cuts through the Fermi level. This allows us to write $G_{3d}({\bf p},\omega)= \mid \phi_{3d}({\bf p})\mid^2[\omega-\epsilon_{\bf p}-i\delta_{\bf p}]^{-1}$ with $\epsilon_{\bf p}$ the eigenenergies, $\phi_{3d}({\bf p})$ the Cu $3d_{x^2-y^2}$ projected character of the band, and $\delta_{\bf p}= \gamma_e$ sign$(\epsilon_{\bf p})$. Expressions for these functions can be determined by Wannier downfolding in density functional based approaches, or via tight-binding parameterizations for free electrons, or exact diagonalization studies for correlated models. In addition expressions for the momentum dependent electron-phonon coupling constants $g_{\nu}^{3d}({\bf p,q})$ can be determined via largely the same methods. 

The localized core Cu $2p$ orbitals are represented by $G_{2p}({\bf p},\omega) = [\omega-E_{2p,3d} + i \Gamma]^{-1}$  having an energy $E_{2p,3d}$ measured between 
the bottom of the Cu $3d$ level and the Cu $2p$ ${L_3}$ level, and a phenomenological core-hole lifetime $1/\Gamma$. One can show that such a resonance approximation for the $2p$ propagator (which neglects the contributions from other edges) works well for RIXS at either the Cu $L_2$ or $L_3$ edge~\cite{Marra2012, Jia2015}.

\begin{widetext}
The limit $\gamma_e \rightarrow 0$ 
leads to the collapse of
the frequency integral. After substitution of the expressions for the Green functions one obtains
\begin{align}
\chi_{\text{bare}}(\mathbf{k_i},\mathbf{k_f},\mathbf{q}|\omega_i,\omega_f,\Omega) = \frac{1}{N} \sum_{\mathbf{p}} \left[\mathbf{e_i}\cdot\mathbf{d}_{3d,2p}(\mathbf{p+q,p-k_f})\right] \left[\mathbf{e_f}\cdot\mathbf{d}_{2p,3d}(\mathbf{p-k_f,p})\right]
\nonumber\\ 
\times\left[\mathbf{e_i}\cdot\mathbf{d}_{2p,3d}(\mathbf{p-k_f,p+q})\right] \left[\mathbf{e_f}\cdot\mathbf{d}_{3d,2p}(\mathbf{p,p-k_f})\right]
\nonumber\\ 
\times \left\{f[-\epsilon_{\mathbf{p+q}}]-f[-\epsilon_{\mathbf{p}}]\right\} \frac{\left|\phi_{3d}(\mathbf{p})\right|^2 \left|\phi_{3d}(\mathbf{p+q})\right|^2}{\left|\epsilon_{\mathbf{p}}-\omega_f-E_{2p-3d}+i\Gamma\right|^2} 
 \delta[\Omega+\epsilon_{\mathbf{p}}-\epsilon_{\mathbf{p+q}}].
\end{align}
By taking the limit $\gamma_e \rightarrow 0$ we neglect the multi-particle interactions within the $3d$ electrons, focusing attention on a simple understanding of the role of lattice coupling. For fully interacting electrons, one still can proceed by taking Eq. (1) in place of (2). 

As the main focus of this study, the phonon contribution (represented diagrammatically in Fig. \ref{fig:Phonon}), can be cast as
\begin{equation}
\chi_{\text{phonon}}(\mathbf{k_i},\mathbf{k_f},\mathbf{q}|\omega_i,\omega_f,\Omega) 
= \frac{1}{2\pi i} \left. \sum_{\nu,\nu'} \Lambda_{\nu}^{(1)}(\mathbf{k_i,k_f,q}|\omega_i,\omega_f,i\zeta) \tilde{D}_{\nu,\nu'}(\mathbf{q},i\zeta) \Lambda_{\nu'}^{(2)}(\mathbf{k_i,k_f,q}|\omega_i,\omega_f,i\zeta) \right|_{i\zeta\to\Omega-i0^+}^{i\zeta\to\Omega+i0^+},
\end{equation}
with $\nu,\nu^{\prime}$ indices running over phonon modes. In accordance with Fermi's golden rule, this contribution involves the product of two matrix elements $\Lambda$ and the phonon density of states represented by the phonon propagator $\tilde{D}$. In the limit of large core-hole energy $E_{2p,3d}$, the matrix elements can be written as 
\begin{align}
\Lambda_{\nu}^{(1)}(\mathbf{k_i,k_f,q}|\omega_i,\omega_f,i\zeta) = \frac{1}{N} \sum_{\mathbf{p}} \left[\mathbf{e_i}\cdot\mathbf{d}_{3d,2p}(\mathbf{p+q,p-k_f})\right] \left[\mathbf{e_f}\cdot\mathbf{d}_{2p,3d}(\mathbf{p-k_f,p})\right] 
\nonumber\\ 
\times g_{\nu}^{3d,3d}(\mathbf{p,p+q}) \int_{-\infty}^{+\infty} d\omega f(-\omega) \left\{ -\frac{1}{\pi} \Im G_{3d}(\mathbf{p+q},\omega)\cdot G_{2p}^{*}(\mathbf{p-k_f},\omega-\omega_i) G_{3d}(\mathbf{p},\omega-i\zeta) \right.
\nonumber\\ 
\left. -\frac{1}{\pi} \Im G_{3d}(\mathbf{p},\omega)\cdot G_{2p}^{*}(\mathbf{p-k_f},\omega-\omega_f) G_{3d}(\mathbf{p+q},\omega+i\zeta) \right\}
\end{align}
and
\begin{align}
\Lambda_{\nu'}^{(2)}(\mathbf{k_i,k_f,q}|\omega_i,\omega_f,i\zeta) = \frac{1}{N} \sum_{\mathbf{p'}} \left[\mathbf{e_i}\cdot\mathbf{d}_{2p,3d}(\mathbf{p'-k_f,p'+q})\right] \left[\mathbf{e_f}\cdot\mathbf{d}_{3d,2p}(\mathbf{p',p'-k_f})\right] 
\nonumber\\ 
\times g_{\nu'}^{3d,3d}(\mathbf{p'+q,p'}) \int_{-\infty}^{+\infty} d\omega' f(-\omega') \left\{ -\frac{1}{\pi} \Im G_{3d}(\mathbf{p'+q},\omega')\cdot G_{2p}(\mathbf{p'-k_f},\omega'-\omega_i) G_{3d}(\mathbf{p'},\omega'-i\zeta) \right.
\nonumber\\ 
\left. -\frac{1}{\pi} \Im G_{3d}(\mathbf{p'},\omega')\cdot G_{2p}(\mathbf{p'-k_f},\omega'-\omega_f) G_{3d}(\mathbf{p'+q},\omega'+i\zeta) \right\}.
\end{align}
Using the previous Green functions definitions these equations become
\begin{align}
\Lambda_{\nu}^{(1)}(\mathbf{k_i,k_f,q}|\omega_i,\omega_f,i\zeta) = \frac{1}{N} \sum_{\mathbf{p}} \left[\mathbf{e_i}\cdot\mathbf{d}_{3d,2p}(\mathbf{p+q,p-k_f})\right] \left[\mathbf{e_f}\cdot\mathbf{d}_{2p,3d}(\mathbf{p-k_f,p})\right] 
\nonumber\\ 
\times g_{\nu}^{3d,3d}(\mathbf{p,p+q}) \frac{\left|\phi_{3d}(\mathbf{p})\right|^2 \left|\phi_{3d}(\mathbf{p+q})\right|^2}{\epsilon_\mathbf{p}-\epsilon_\mathbf{p+q}+i\zeta+i\gamma_e} 
\nonumber\\ 
\times \left[\frac{f(-\epsilon_\mathbf{p})}{\epsilon_\mathbf{p}-\omega_f-E_{2p,3d}-i\Gamma} - \frac{f(-\epsilon_\mathbf{p+q})}{\epsilon_\mathbf{p+q}-\omega_i-E_{2p,3d}-i\Gamma}\right]
\end{align}
and
\begin{align}
\Lambda_{\nu'}^{(2)}(\mathbf{k_i,k_f,q}|\omega_i,\omega_f,i\zeta) = \frac{1}{N} \sum_{\mathbf{p'}} \left[\mathbf{e_i}\cdot\mathbf{d}_{2p,3d}(\mathbf{p'-k_f,p'+q})\right] \left[\mathbf{e_f}\cdot\mathbf{d}_{3d,2p}(\mathbf{p',p'-k_f})\right] 
\nonumber\\ 
\times g_{\nu'}^{3d,3d}(\mathbf{p'+q,p'}) \frac{\left|\phi_{3d}(\mathbf{p'})\right|^2 \left|\phi_{3d}(\mathbf{p'+q})\right|^2}{\epsilon_\mathbf{p'}-\epsilon_\mathbf{p'+q}+i\zeta+i\gamma_e} 
\nonumber\\ 
\times \left[\frac{f(-\epsilon_\mathbf{p'})}{\epsilon_\mathbf{p'}-\omega_f-E_{2p,3d}+i\Gamma} - \frac{f(-\epsilon_\mathbf{p'+q})}{\epsilon_\mathbf{p'+q}-\omega_i-E_{2p,3d}+i\Gamma}\right].
\end{align}
Here we do not assume free electrons, and keep $\gamma_e$ finite.

These matrix elements carry the resonant enhancement of phonon scattering via coupling to electrons. Importantly, they show that the resonance process itself imparts a projection of the electron-phonon coupling to the portion of the coupling involving the photo-excited valence electron, in this case copper. If we instead focused on the oxygen $K$-edge, the matrix elements would involve the projection onto the oxygen contribution to the hybridized valence band. This is one important facet that already distinguishes RIXS from other measurements, and allows RIXS from phonons to have an element selectivity to not only the particle-hole excitations, but also the momentum-dependent phonon coupling to them.

Finally, for the phonon contribution one obtains
\begin{align}
&\chi_{\text{phonon}}(\mathbf{k_i},\mathbf{k_f},\mathbf{q}|\omega_i,\omega_f,\Omega) = \sum_{\nu,\nu'} \frac{1}{N} \sum_{\mathbf{p}} \frac{1}{N} \sum_{\mathbf{p'}}
\nonumber\\ 
& \times\left[\mathbf{e_i}\cdot\mathbf{d}_{3d,2p}(\mathbf{p+q,p-k_f})\right] \left[\mathbf{e_f}\cdot\mathbf{d}_{2p,3d}(\mathbf{p-k_f,p})\right] g_{\nu}^{3d,3d}(\mathbf{p,p+q}) 
\nonumber\\ 
&\times \left|\phi_{3d}(\mathbf{p})\right|^2 \left|\phi_{3d}(\mathbf{p+q})\right|^2
 \left[\frac{f(-\epsilon_\mathbf{p})}{\epsilon_\mathbf{p}-\omega_f-E_{2p,3d}-i\Gamma} - \frac{f(-\epsilon_\mathbf{p+q})}{\epsilon_\mathbf{p+q}-\omega_i-E_{2p,3d}-i\Gamma}\right]
\nonumber\\ 
&\times \left[\mathbf{e_i}\cdot\mathbf{d}_{2p,3d}(\mathbf{p'-k_f,p'+q})\right] \left[\mathbf{e_f}\cdot\mathbf{d}_{3d,2p}(\mathbf{p',p'-k_f})\right] g_{\nu'}^{3d,3d}(\mathbf{p'+q,p'})
\nonumber\\ 
&\times \left|\phi_{3d}(\mathbf{p'})\right|^2 \left|\phi_{3d}(\mathbf{p'+q})\right|^2 
\left[\frac{f(-\epsilon_\mathbf{p'})}{\epsilon_\mathbf{p'}-\omega_f-E_{2p,3d}+i\Gamma} - \frac{f(-\epsilon_\mathbf{p'+q})}{\epsilon_\mathbf{p'+q}-\omega_i-E_{2p,3d}+i\Gamma}\right]
\nonumber\\ 
&\times\frac{1}{\pi} \Im \left\{\frac{1}{\epsilon_\mathbf{p}-\epsilon_\mathbf{p+q}+\Omega+i\gamma_e} \tilde{D}_{\nu,\nu'}(\mathbf{q},\Omega) \frac{1}{\epsilon_\mathbf{p'}-\epsilon_\mathbf{p'+q}+\Omega+i\gamma_e}\right\}.
\end{align}

It can easily be seen that there can be several sources of enhancement from the resonant process. First, if both the ${\bf p}$
and ${\bf p+q}$ states are occupied, no scattering can occur since a core electron is Pauli blocked. Therefore at least one or both of those states must be unoccupied. However the largest contribution results when only one of the momentum states is unoccupied so that an occupied electron state can refill the core hole.

Since we have assumed that the phonons are eigenstates indexed by $\nu$,
the loss peaks are determined by the phonon mode energies and their dispersion with momentum, multiplied by overall factors that represent the resonant excitation process. This demonstrates how the electron-hole pairs created by the process couple to those phonons in a symmetry-dependent way that is sampled by the incident and scattered polarization orientations. This is analogous to the way in which photon polarization selection rules can be used to probe unconventional superconductivity \cite{RamanRMP,tpd}.
\end{widetext}

\section{General behavior of RIXS from Phonons}

We can examine a few insights from the momentum-dependent electron-phonon coupling matrix element. From Eqs (6-8), it is clear that the overall intensity of the phonon contribution to RIXS is governed by the strength and momentum dependence of the electron-phonon coupling $g_{\nu}^{3d,3d}({\bf q})$. We first consider general couplings that determine its momentum dependence.

For electron-phonon coupling of the deformation type, which involves lattice motion that modifies the kinetic energy of electrons via orbital overlaps on adjacent ions, the electron-phonon coupling depends strongly on transfered momentum ${\bf q}$, being largest for large momentum transfers and vanishing in the limit of ${\bf q}\rightarrow 0$ \cite{Mahan}. These phonons, such as the longitudinal acoustic modes or the bond-stretching modes in the cuprates, fail to appear in Raman or optical conductivity and likewise would not contribute to RIXS for small momentum transfers via Eq. (6-7). Moreover, the largest RIXS intensity from those phonons will occur for large momentum at the BZ boundaries where deformation electron-phonon coupling is largest. 

On the other hand, for couplings that are driven by modulations of the electrostatic energies via atomic motion, the momentum dependence is biased towards small momentum transfers. These include out-of-plane $c$-axis oxygen vibrations in transition metal oxides, which show up in Raman and optical measurements and likewise in RIXS at small wavevector transfers \cite{Johnston}.

Also any phonons which do not transform according the full irreducible point group symmetry of the lattice will likewise vanish at zone center {provided that the dipole matrix elements do not depend upon the magnitude of the momenta involved in scattering. Since the dipole matrix element involves photo excitations of core levels and is thus quite local, this would be a very good approximation. As a consequence, scattering would be suppressed from zone center phonons having $B_{1,2}$ or $E_g$ symmetries in $D^{4h}$ systems. This is quite different from resonant optical Raman scattering where electrons can be non-locally excited out of the valence band into fully itinerant bands and thus depend on the momenta involved in the scattering process.

We can show that this is the case for Cu $L-$edge RIXS via the following consideration. Following Ref. \onlinecite{Ament}
the RIXS vertex in the dipole approximation (here for the process from the initial to intermediate state of RIXS) is
\begin{align} 
\hat{V} = \frac{1}{\sqrt{N}} \sum_{j=1}^N e^{i {\bf k}_{\rm in} \cdot {\bf j}} 
\hat{{\bf r}}_{\bf j} \cdot {\bf e}_{\rm in}, 
\end{align}
where ${\bf k}_{\rm in}$ (${\bf e}_{\rm in}$) is the incoming photon momentum (polarization), ${\bf j}$ is the lattice site, 
and $\hat{{\bf r}}_{\bf j} $ is the position operator of electron at site ${\bf j}$.

\begin{widetext}
This can be written in the second quantization form (for the transition metal ion $L$ edge) as
\begin{align}
\hat{V} = &\sum_{\bf q, \alpha, \beta} ({\bf d}_{2p \alpha, 3d \beta}   c^\dag_{{\bf q}+{\bf k}_{\rm in}, 2p \alpha} c_{{\bf q}, 3d \beta} + {\bf d}_{2p \alpha, 3d \beta} c^\dag_{{\bf q}+{\bf k}_{\rm in}, 3d \beta } c_{{\bf q}, 2p \alpha }) {\bf e}_{\rm in},
\end{align}
where $\alpha$ ($\beta$) are the quantum numbers of the electron in the $2p$ ($3d$) orbital.
\end{widetext}
Here the RIXS dipole operator matrix element ${\bf d}_{2p \alpha, 3d \beta}$  does not depend on the momentum of the valence or core electrons,
since it follows from the expression
\begin{align}
\hat{{\bf r}}_{\bf j} = \sum_{n, n', {\bf l}, {\bf m}} c^\dag_{{\bf j} n} | \phi_{{\bf l} n} \rangle \langle \phi_{{\bf l} n} | \hat{{\bf r}}_{\bf j}  | \phi_{{\bf m} n'} \rangle \langle \phi_{{\bf m} n'}|  c_{{\bf j} n} 
\end{align}
(where $n$ and $n'$ are quantum numbers of the electrons in the $3d$ or $2p$ orbitals, ${\bf l, m}$ are lattice sites, and 
$\mid\phi_{{\bf j}, n}\rangle$ are Wannier states centered on atom ${\bf j}$)
and the fact that   
 \begin{align}
\langle \phi_{{\bf l} 2p \alpha} | \hat{{\bf r}}_{\bf j}  | \phi_{{\bf m} 3d } \rangle \simeq
 \delta_{{\bf m j}}\delta_{{\bf l j}} \times {\bf f}(\alpha, \beta) \equiv  {{\bf d}_{2p \alpha, 3d \beta}}.
 \end{align}
This is due to the localized nature of the core hole. This property is 
the one that distinguishes RIXS from optical Raman scattering.

In this paper where we are interested only in RIXS in the energy range of single phonons, we keep only the $\beta=d_{x^2-y^2}$ orbital and neglect the other Cu 3$d$ orbitals which would appear if we consider larger energy transfers at the $d-d$ excitation lines. Moreover, we are solely interested in the RIXS transitions for which the spin of the $3d$ electrons is conserved: this is possible when choosing a particular combination of the polarization vectors ${\bf e_{i}}$ and  ${\bf e_{f}}$ ~\cite{Groot1998, Veenendaal2006, AmentPolarization, Haverkort2010, Marra2012}. Thus, assuming that the latter indeed takes place, we can neglect the spin quantum number carried by the $3d$ electron and we are left with the dipole matrix elements which depend solely on the $2p$ orbital quantum number: ${{\bf d}_{2p \alpha, 3d \beta}} \equiv {{\bf d}_{2p \alpha, 3d}}$ where $\alpha=x, y$. We note that in RIXS experiment for cuprates, different momentum transfers are usually realized by changing the relative angle between the incoming photons and the sample. Thus, the incoming and outgoing photon polarizations ${\bf e_{i,f}}$ may effectively depend on the momentum transfer~\cite{AmentPolarization}. We do not consider this effect in the current manuscript. 


\section{An example: 8-Band C\lowercase{u}-O model}
We can evaluate RIXS for a simple model as given in Refs. \onlinecite{Johnston} and \onlinecite{SJJacoustic} which derived momentum dependent electron-phonon couplings for Cu-O planar longitudinal acoustic, bond stretching and bond bending modes, as well as $c-$axis apical modes. Besides accounting for the phonon eigenmodes, the model consists of 2 Cu (3$d_{x^2-y^2}$ and $4s$) orbitals octahedrally coordinated and hybridized with O $2p_{x,y}$ orbitals that are $\sigma$-bonded to the Cu 3$d_{x^2-y^2}$ orbital, plus the apical O $2p_z$ orbital hybridized with Cu $4s$. Oxygen hopping is taken into account.
We finally add the 3 Cu $2p$ core levels from which electrons are resonantly excited.

For the phonon excitations, we specifically focus on the oxygen modes common to all perovskite oxides. These phonons typically couple to CuO electrons either electrostatically in the case of $c-$axis phonon modes or via bond deformation for in-plane stretching modes. The forms for the electron-phonon coupling depend strongly on which process is dominant, and further can depend strongly on either fermion momentum ${\bf k}$ or phonon momentum transfer ${\bf q}$. In Ref. \onlinecite{Johnston} these modes were examined and expressions for their couplings were derived from general symmetry grounds. Their forms can be expressed as:
\begin{eqnarray}
g_{LA,br}({\bf k,q}) = &&g_{LA,br}^{0}\sqrt{[\sin^2(q_x/2)+\sin^2(q_y/2)]}/2,\nonumber \\
g_{A1,B1} ({\bf k,q}) = &&g^{0}_{A1,B1}[\sin(k_x/2)\sin(p_x/2)\cos(q_y/2)\nonumber\\
\pm&&\sin(k_y/2)\sin(p_y/2)\cos(q_x/2)]/2, \nonumber\\
g_{apex}({\bf k,q})=&&g^0_{apex}[\cos(k_x/2)-\cos(k_y/2)]\nonumber\\
\times &&[\cos(p_x/2)-\cos(p_y/2)]/4,
\end{eqnarray}
with ${\bf p=k+q}$ and the lattice constant $a$ set to 1. Here $LA$ denotes the longitudinal acoustic modes, $br$ the Cu-O in-plane bond stretching modes, $A1,B1$ the $c$-axis oxygen modes involving charge transfer between Cu-O and O-O, respectively, and $apex$ the apical oxygen modes. The prefactors $g^0_{\nu}$ for mode $\nu$ are rather weakly dependent on momenta and are generally set by the orbital content of the bands that are coupled to the lattice modes. Here we limit focus on them to solely setting an overall magnitude relative to each other. For more details, the reader is directed to Ref. \onlinecite{Johnston}. Finally we take the apical and $A1,B1$ modes to be dispersionless phonons of frequencies $\Omega=$ 85, 40 and 35 meV, respectively, and the dispersions of the acoustic and breathing modes in eV as $\Omega_{LA}=0.015\sqrt{[\sin^2(q_x/2)+\sin^2(q_y/2)]/2}$ and $\Omega_{br}=0.085[1-0.18\sqrt{\sin^2(q_x/2)+\sin^2(q_y/2)}]$ to match the dispersions in Bi-based bi-layer cuprates \cite{BiPhonons}.

We neglect any momentum or angle dependence of the 
polarization vectors 
and take them to be constants such that Eqs. (1) and (8) can be simply integrated numerically.  Here we have taken a simple downfolded three-band Cu-O tight-binding model with Cu-O hopping $t_{pd}=1.6$eV and charge transfer 0.9 eV  to mimic the anti-bonding band structure $\epsilon_{\bf k}$ of the cuprates, doped for an overall filling $\langle n\rangle\sim0.84$ \cite{neutron,buckling}. Additionally we have set $\omega_i+E_{2p,3d}=0$ since the resonance is quite broad due to the short core-hole lifetime ($\Gamma=0.5$ eV). The width of the all phonons is set to 5 meV, and $\gamma_e=0.010$eV$+\omega$ is taken as the width of the $\delta$-function in Eqs. (1) and (8) to phenomenologically mimic inelastic relaxation in the particle-hole continuum. The resulting RIXS response is shown in the Figs. \ref{fig:RIXS0}-\ref{fig:RIXS5} for momentum cuts along the BZ $x$-axis. Since the phonons enter additively, we plot them individually as their intensities can be set from considerations given in the references cited above.

\begin{figure}
\includegraphics[width=\columnwidth]{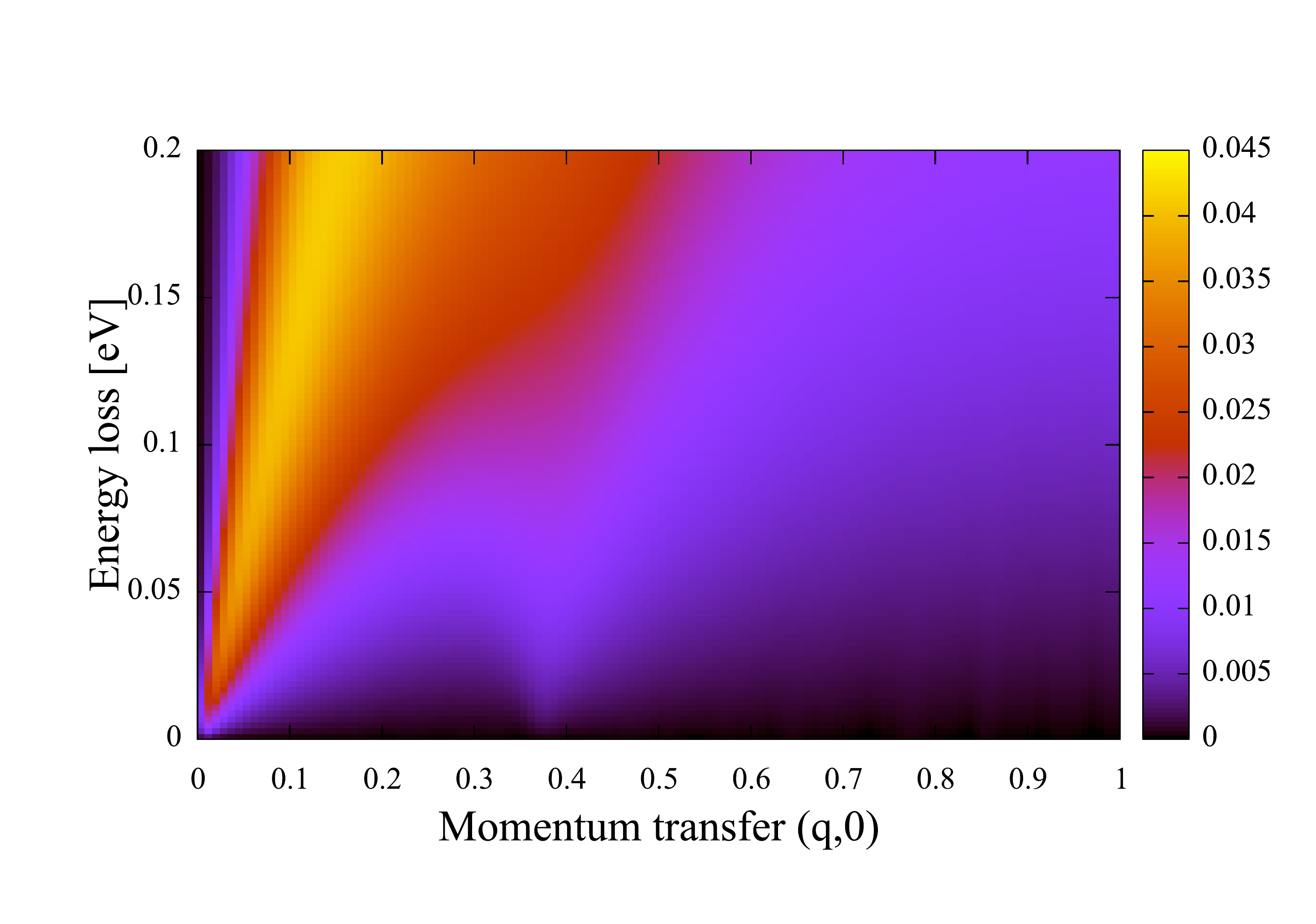}
\caption{\label{fig:RIXS0}  
{Bare contribution to RIXS from particle-hole scattering. Momentum transfer is given in units of $\pi/a$.}}
\end{figure}
Fig. \ref{fig:RIXS0} shows the bare contribution from particle-hole excitations to RIXS from Eq. (1). The particle-hole continuum is resonantly enhanced, with the leading edge rapidly dispersing away from BZ center with the Fermi velocity. The overall continuum expands with increasing energy loss and momentum transfer due to kinematic conditions of particle-hole creation, with the overall strength and sharpness of the continuum being set by the inelastic relaxation $\gamma_e$. We note that the continuum dips down close to zero energy transfer around a momentum $(q,0)\sim 0.4\pi$, reflecting the weak nesting condition of the band structure and Fermi surface for this momentum transfer.

\begin{figure}
\includegraphics[width=\columnwidth]{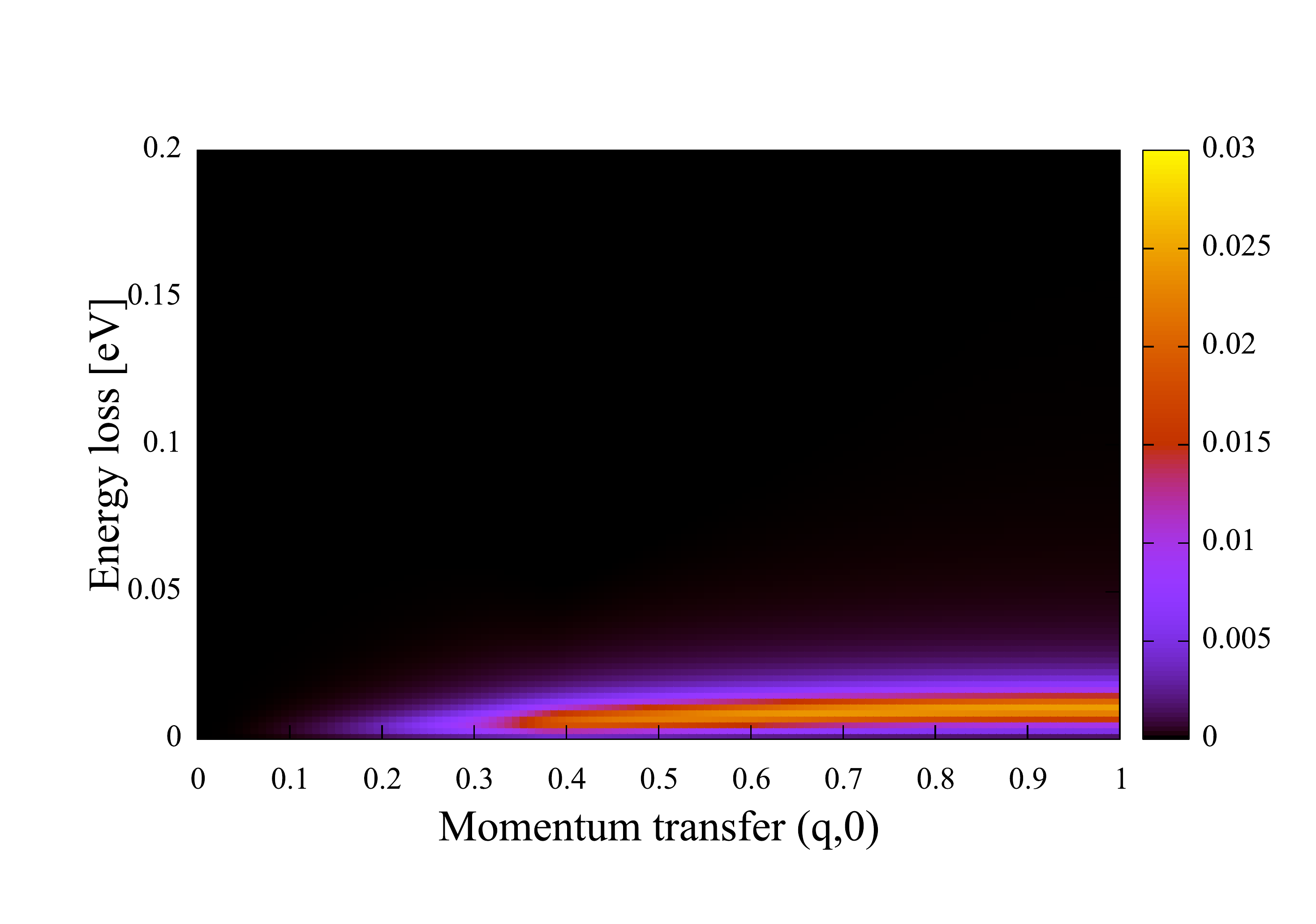}
\caption{\label{fig:RIXS}  
{Longitudinal acoustic phonon contribution to RIXS. Momentum transfer is given in units of $\pi/a$.}}
\end{figure}

Figs. \ref{fig:RIXS}-\ref{fig:RIXS4} show the single phonon contribution to RIXS for the individual phonons listed above. For the acoustic phonon branch (Fig. \ref{fig:RIXS}) it is clear that the dispersion for RIXS follows the phonon dispersion, having an intensity which follows the momentum dependence of the electron-phonon coupling $\sim\sin^2(qa/2)$. The relative strength of the phonon contribution to RIXS compared to the bare response (Fig. \ref{fig:RIXS}) is set both by the overall electron-phonon coupling as well as the particle-hole relaxation $\gamma_e$. Therefore while the overall magnitude of the electron-phonon coupling cannot be straightforwardly determined solely from the single phonon RIXS signal, the momentum dependence of the coupling as well as the dispersion can be clearly read off directly from the RIXS intensity plots. In contrast, the magnitude of the coupling can be determine by the ratio of the phonon side bands when multiple phonon inelastic events are considered \cite{LeePhononsI,LeePhononsII}.

\begin{figure}
\includegraphics[width=\columnwidth]{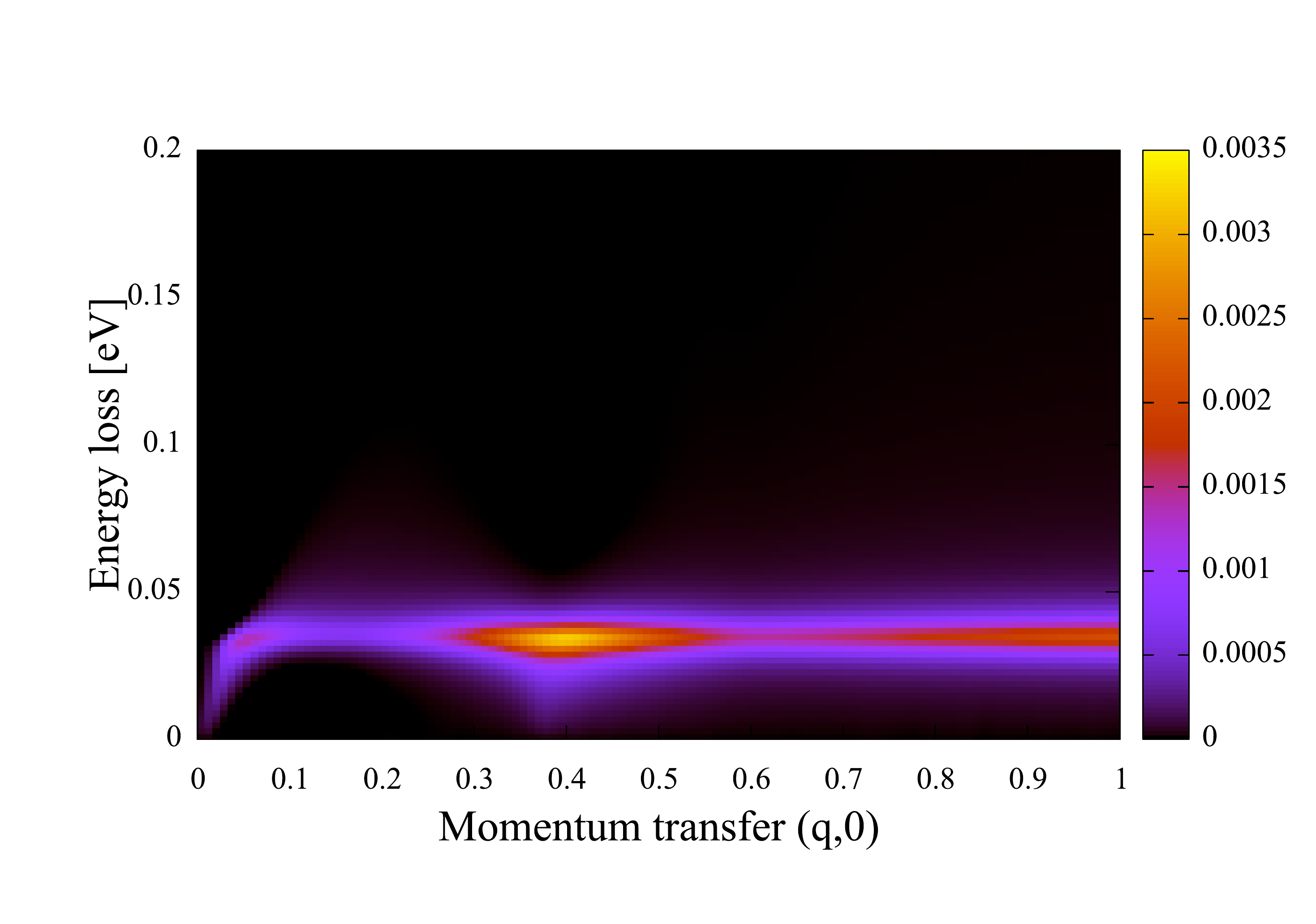}
\caption{\label{fig:RIXS1}  
{$B1$ phonon contribution to RIXS. Momentum transfer is given in units of $\pi/a$.}}
\end{figure}

Consideration of the interplay of electron-phonon coupling to the continuum as well as the overall symmetry between the electron-phonon coupling and the $L-$edge RIXS process can be seen clearly in Fig. \ref{fig:RIXS1} which shows the contribution to RIXS from the $B1$ phonons (charge transfer between O$_x$ and O$_y$ in the Cu-O unit cell). Both the overall dispersion as well as the intensity vary nonmonotonically across the BZ. We first consider symmetry effects. As already noted, this phonon branch at zone center transforms as a lower representation of the $D^{4h}$ point group, and therefore, due to the locality of the RIXS process, the $B1$ phonon branch does not contribute to the RIXS intensity at zone center and grows for increasing momentum transfer. This is in strong contrast to either resonant optical Raman scattering or infrared conductivity for which the intensities are strong for this phonon and show clear Fano interference with the particle-hole continuum \cite{buckling}.

Moreover, a clear interference effect is seen when the particle-hole continuum crosses the phonon line. This occurs weakly near zone center and much more clearly around the weak nesting momentum $q \sim 0.4\pi/a$. At both of these points the phonon contribution to RIXS inherits contributions from the particle-hole continuum and shows a strong momentum-dependent Fano lineshape, distorting along the lines of the soft continuum. The strength of this Fano effect is like that in Raman scattering, determined both by the strength of the electron-hole continuum at the frequency of the phonon and the magnitude of electron-phonon coupling, but in addition the range of momentum transfers available through RIXS gives a clear qualitative way of understand electron-lattice coupling nearinstabilities, allowing for the particle-hole continuum to interfere with the phonon in a narrow range of mode momenta.

\begin{figure}
\includegraphics[width=\columnwidth]{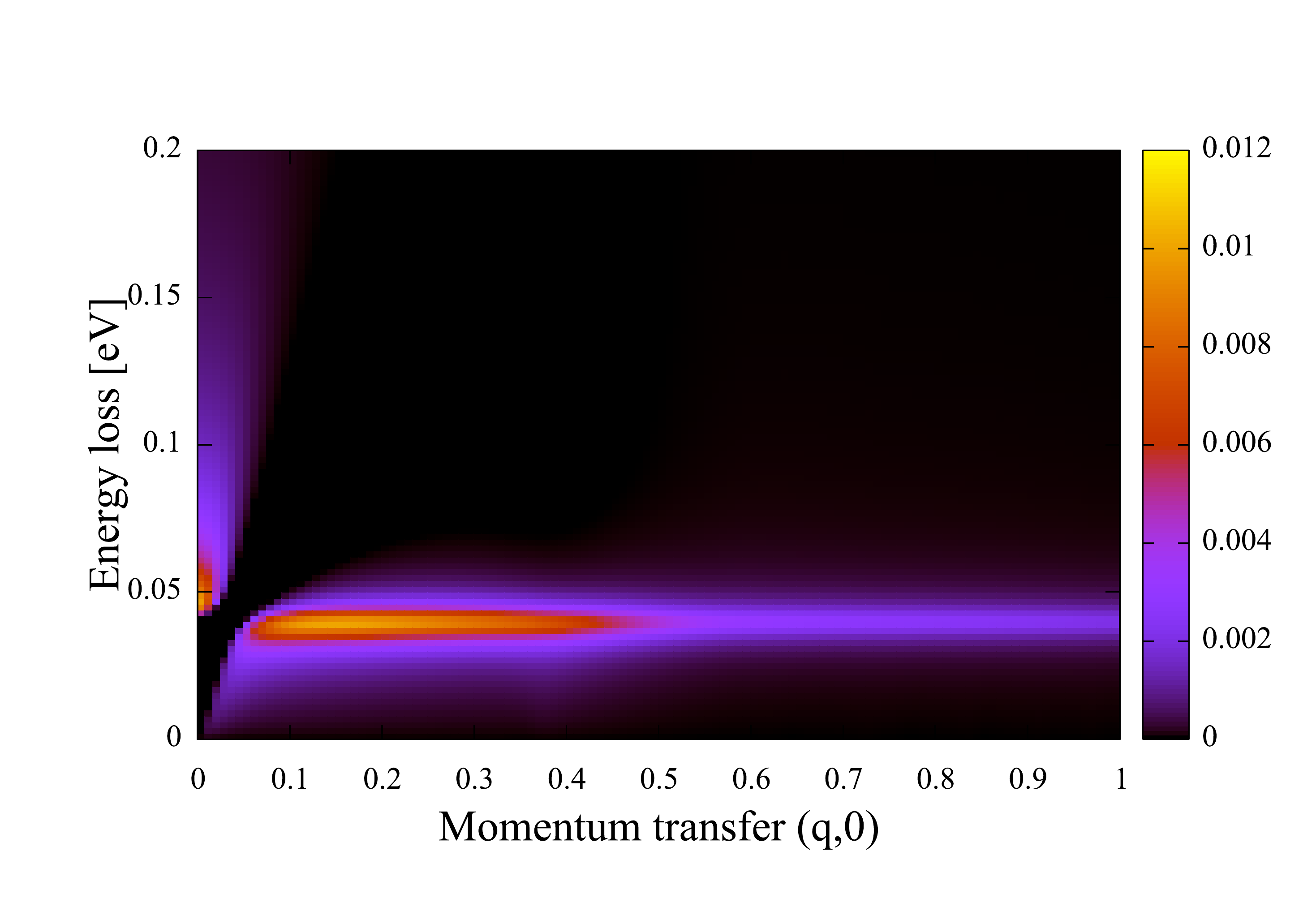}
\caption{\label{fig:RIXS2}  
{A$_1$ phonon contribution to RIXS. Momentum transfer is given in units of $\pi/a$.}}
\end{figure}

This Fano effect is indicated also clearly in the $A1$ phonon contribution to RIXS, shown in Fig. \ref{fig:RIXS2}. Here the symmetry of the $A1$ modes (involving charge transfer between Cu-O) allow for finite coupling at zone center, where the overall bare electron-phonon coupling is largest. However a clear drop in intensity is observed where the electron-hole continuum crosses the phonon energy. This intensity would add to the intensity of the bare continuum shown in Fig. \ref{fig:RIXS0}. The momentum-dependent coupling falls off for large momentum transfers and gives vanishing intensity at the BZ boundary.

\begin{figure}
\includegraphics[width=\columnwidth]{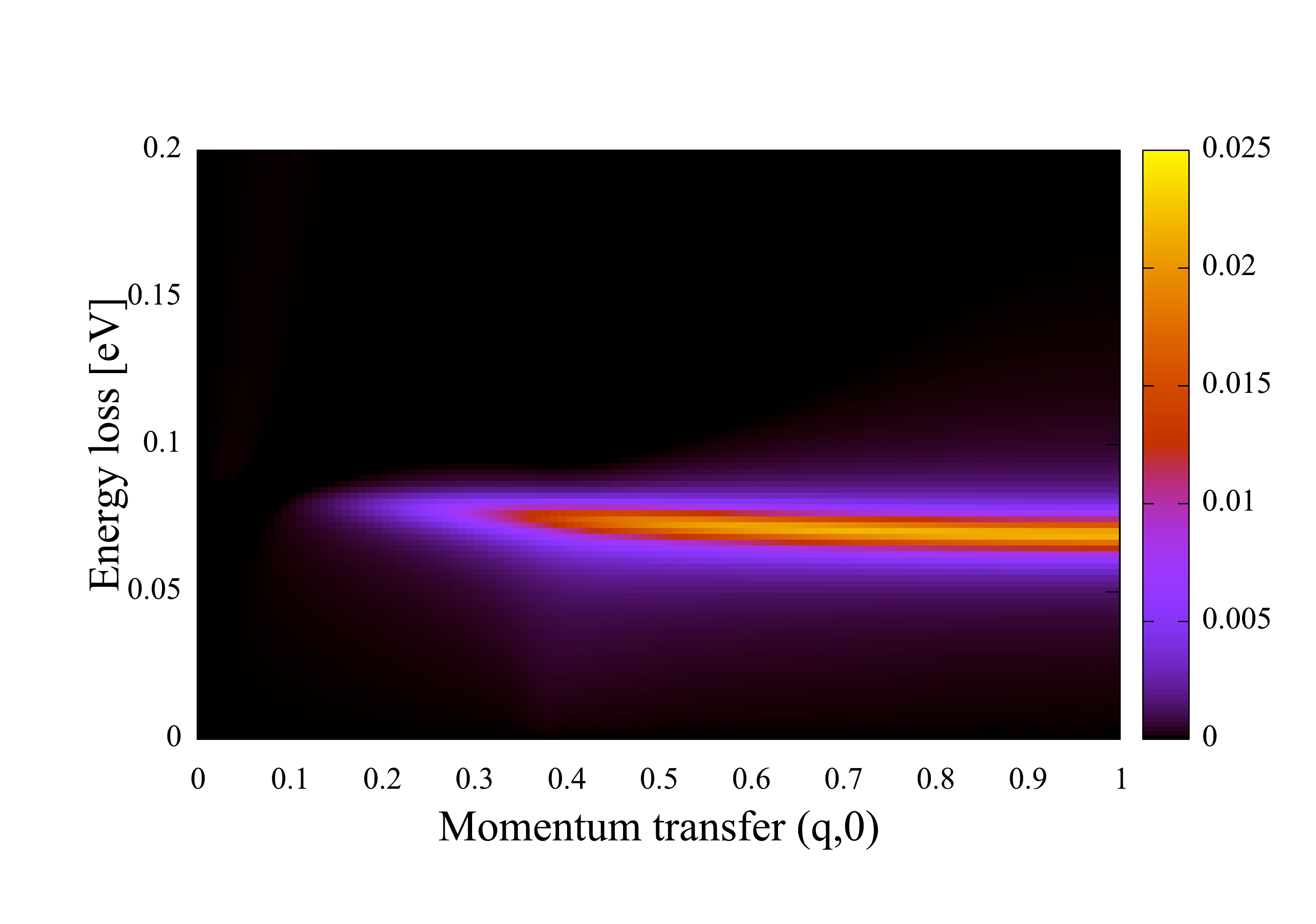}
\caption{\label{fig:RIXS3}  
{Cu-O bond-stretching (breathing) phonon contribution to RIXS. Momentum transfer is given in units of $\pi/a$.}}
\end{figure}

In Fig. \ref{fig:RIXS3}, the Cu-O bond stretching phonons disperse downward away from zone center, and due to the strong momentum dependence of the bare electron-phonon coupling, the intensity grows as $\sin^2(qa/2)$ to be largest at the BZ boundary. The coupling's momentum dependence minimizes the Fano effect near the BZ zone center, yet allows for a weak coupling around the weak nesting feature near $q\sim0.4\pi/a$ where the intensity extends from the phonon line down to lower energy transfers.

\begin{figure}
\includegraphics[width=\columnwidth]{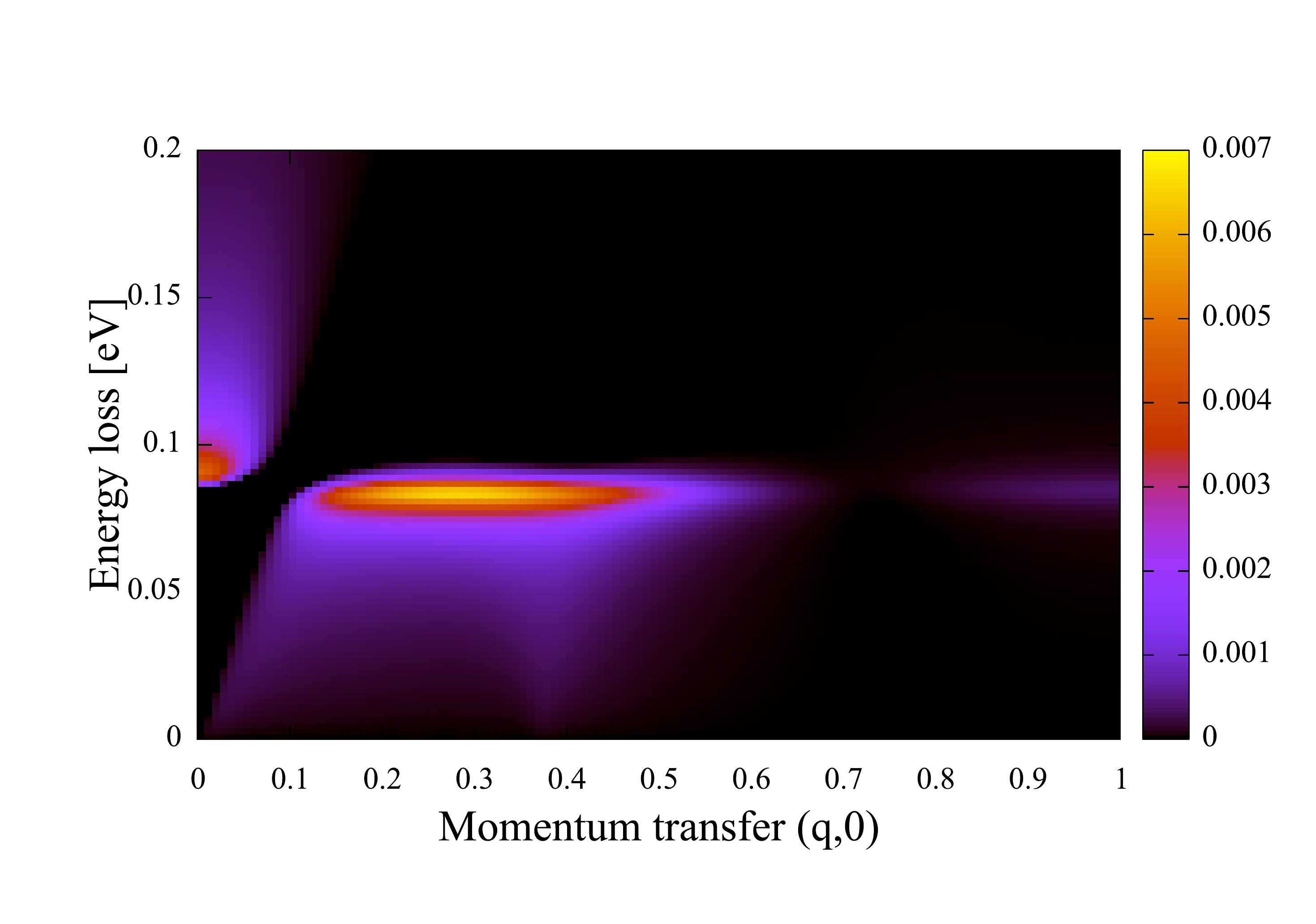}
\caption{\label{fig:RIXS4}  
{Apical oxygen phonon contribution to RIXS. Momentum transfer is given in units of $\pi/a$.}}
\end{figure}

In Fig. \ref{fig:RIXS4} the dispersionless apical oxygen contribution to RIXS is shown. The bare electron-phonon coupling is quite anisotropic in momentum space due to the charge transfer pathway between planar Cu and the apical oxygens, and reaches its strongest coupling for zone center. However, as in the case of $A1$ phonons, the Fano effect is clearly observed at the intersection of the apical phonon with the main peak in the electron-hole continuum. The bare intensity would normally fall smoothly for larger momentum transfers, yet in contrast to the $A1$ phonons and more like the $B1$ phonons,  the intensity remains large as the phonon line disperses across the weak nesting momentum before abruptly falling for larger momentum transfers.

\begin{figure}
\includegraphics[width=\columnwidth]{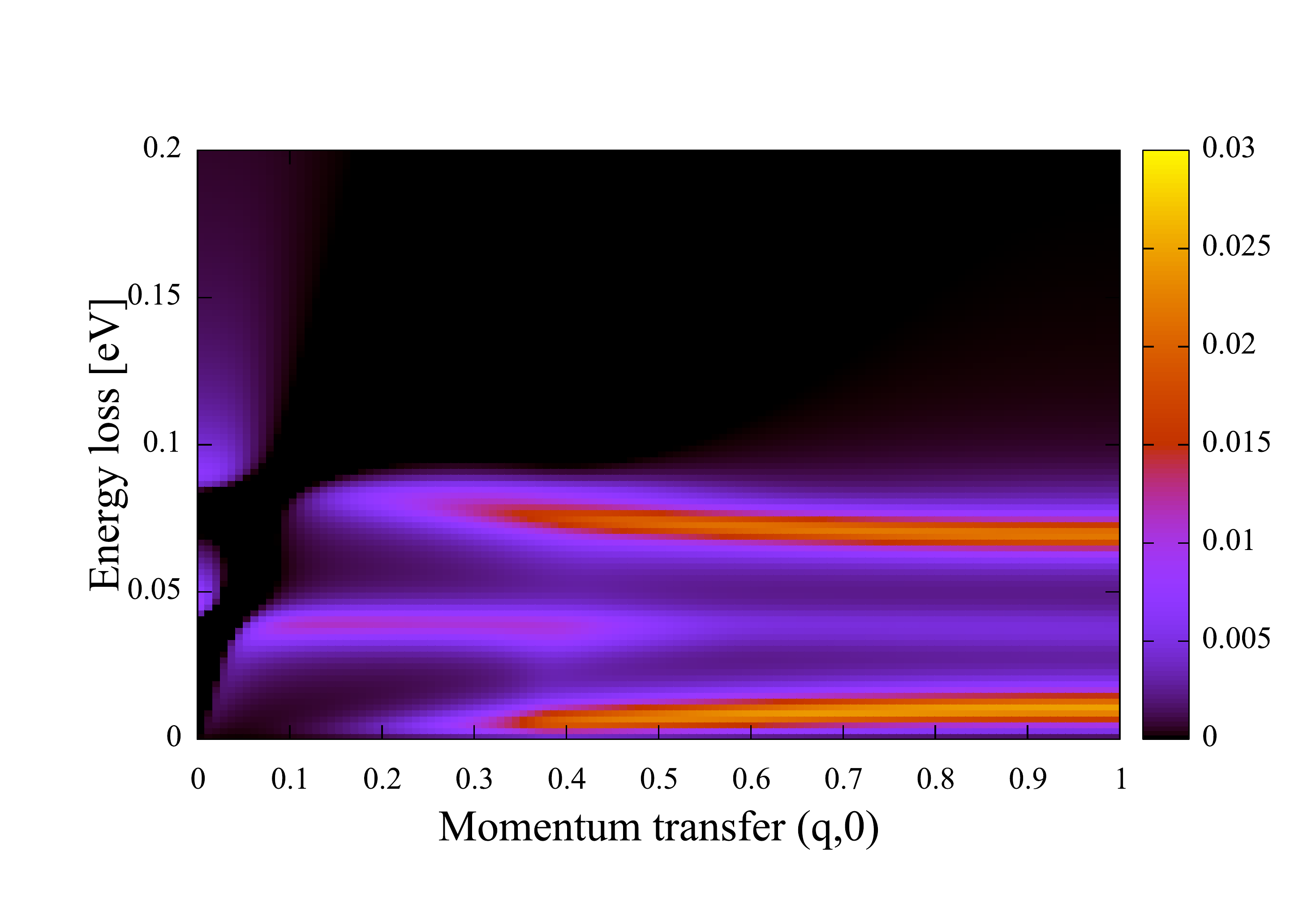}
\caption{\label{fig:RIXS5}  
{Sum of all contributions from phonons to RIXS. Momentum transfer is given in units of $\pi/a$.}}
\end{figure}

As the electron-phonon coupling is only considered at lowest order, these contributions add to the overall RIXS signal, as shown in Fig. \ref{fig:RIXS5}. One can clearly see that different phonon contributions are illuminated depending upon the overall magnitude of the electron-phonon coupling as well as the intersection of the phonon line with the electron-hole continuum.

\begin{figure}
\includegraphics[width=\columnwidth]{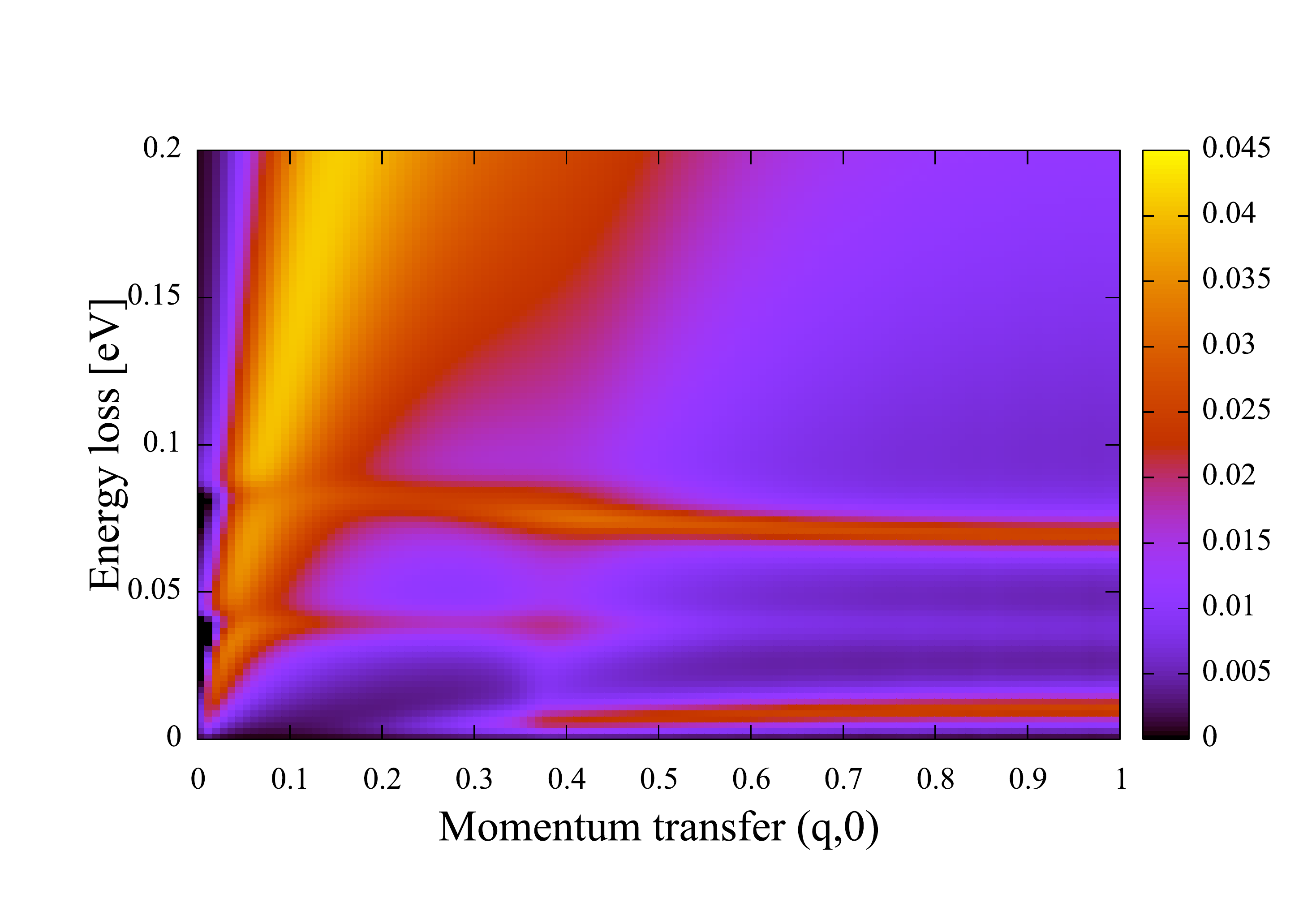}
\caption{\label{fig:RIXS6}  
{Sum of all contributions. Momentum transfer is given in units of $\pi/a$.}}
\end{figure}

Finally, Fig. \ref{fig:RIXS6} plots the total RIXS intensity map from both the bare electron-hole continuum and the contributions from all phonon branches considered. The figure demonstrates the rich overall spectra having a strong momentum dependence. A detailed examination of the structure of the RIXS maps offers an unique perspective into the characteristics of lattice-electron coupling in a way which complements neutron and non-resonant x-ray scattering, as well as Raman and infrared conductivity. Thus RIXS at low energies offers a direct insight into couplings to lattice modes.

We add a remark concerning the effect of screening via the long-range Coulomb interaction, which has not been included in these considerations. It is well known that Coulomb screening mediates charge backflow that enforces particle number conservation and can lead to strong modifications of the electron-phonon coupling for small momentum transfers provided the symmetry of the coupling coincides with the full symmetry of the lattice \cite{tpd,multiband,RamanRMP}. In addition in layer materials a screening enhancement can occur when the phonon energy crosses the effective 3D plasmon \cite{Johnston}. Both of these effects become prominent for small momentum transfers and do not effect the overall considerations given here for larger momentum transfers.

\section{Summary and Connection to experiments}

With the recent development of x-ray Synchrotron Light Source and facilities, two significant improvements have made RIXS a possible approach for directly addressing phonon excitations. First, RIXS for the transition-metal $L$-edge has shown energy resolution around 100meV, enabling RIXS to study not only the specific phonon modes on top of the $d-d$ excitations or at the oxygen $K$-edge directly off the elastic line, but also the momentum dependent one-phonon contributions from different phonon modes directly. Second, although previously polarization discrimination had been limited to only the incoming photon, the recently constructed ESRF RIXS facility has made fully distinguishable polarization measurement for both incoming and outgoing photons possible \cite{LeTaconPolarization}. Other end-stations also would allow for the full photon polarization discrimination in the near future. Full  polarization discrimination measurements can disentangle spin-flip, or single (para)magnon, excitations when the incoming and outgoing photon polarizations are perpendicular to one another (or cross-polarization) and the non-spin-flip or electron-hole excitations when the incoming and outgoing photon polarizations are parallel with each other (or parallel-polarization). 

With the recent development of x-ray Synchrotron Light
Source and facilities, two significant improvements have
made RIXS a possible approach for directly approaching of
the phonon excitations. First, RIXS for the transition-metal $L$-
edge has reached at the ESRF an energy resolution down to 35 meV in the most difficult case (Cu-$L$) among $3d$, enabling RIXS to study not only the specific phonon modes on top of
the $d-d$ excitations or at the oxygen K-edge, but also the
momentum dependent one-phonon contributions from different
phonon modes directly \cite{esrf}. Second, although previously the
phonon polarization discrimination is only for the incoming
photon, recently constructed ESRF RIXS facility has made
the fully distinguishable polarization measurement for both
incoming and outgoing photons possible \cite{Ghiringhelli2006}  with the first application \cite{LeTaconPolarization}.  With the polarimeter  the present standard is 90 meV  with a reduction of the counting rate by a factor of 10.


To summarize, we have derived and calculated the electron-hole and one-phonon contributions to direct RIXS, specifically at the Cu $L$-edge, based on a simple diagrammatic approach. We have examined an 8-band Cu-O model with oxygen phonon modes, including Cu-O planar longitudinal acoustic, Cu-O bond-stretching and apical O modes. This model reveals that the momentum dependence of the electron-phonon coupling can be extracted from the intensity of RIXS excitations in the low energy regime, touting RIXS as a potential method to directly characterize the momentum dependent electron-phonon coupling. Our theoretical approach provides an intuitive picture of how different phonon modes contribute to RIXS and interacts with charge densities, making RIXS a potential new technique for electron-phonon coupling measurement, complementary to ARPES, inelastic neutron scattering and non-resonant inelastic x-ray scattering. 

\acknowledgements

This research was supported by the U.S. Department of Energy (DOE), Office of Basic Energy Sciences, Division of Materials Sciences and Engineering, under Contract No. DE-AC02-76SF00515, SLAC National Accelerator Laboratory (SLAC), Stanford Institute for Materials and Energy Sciences.
K.W. acknowledges support from the Polish National Science Center (NCN) under Project No. 2012/04/A/ST3/00331.

\end{document}